# VIBRATING WIRE FOR BEAM PROFILE SCANNING

Arutunian S.G., Dobrovolski N.M.,
Mailian M.R., Sinenko I.G., Vasiniuk I.E.

Yerevan Physics Institute, Br. Alikhanian St. 2, 375062 Yerevan, Armenia
e-mail: femto@uniphi.yerphi.am

## Abstract

The method for measurement of transverse profile (emittance) of the bunch by detecting of radiation arising at the scattering of the bunch on scanning wire is wide-spread. In this work the information about bunch scattering is proposed to obtain by measuring the oscillation frequency of the tightened scanning wire. In such way the system of radiation (or secondary particles) extraction and measurement can be removed. The entire unit consist of a compact fork with tightened wire and a scanning system.

Normal oscillations frequency of a wire depends on wire tension, its geometric parameters and, in second approximation, on its elastic characteristics. Normal oscillations are generated due to interaction of an alternating current through the wire with magnetic field of a permanent magnet. In this case the magnetic field of the accelerator (field of dipole magnets or quadrupole magnets) is suggested to use for excitation of oscillations. Dependence of oscillations frequency on beam scattering is determined by several factors, including changes of wire tension caused by transverse force of the beam, influence of beam self field. Preliminary calculations show that influence caused by wire heating will dominate.

We have studied strain gauges on the basis of vibrating wire from various materials (tungsten, beryl bronze, niobium zirconium alloys). A scheme of normal oscillations generation by alternating current in autogeneration circuit with automatic frequency adjustment was selected. Special method of wire fixation and elimination of transverse degrees of freedom allow to achieve relative stability better than $10^{-5}$ during several days. For a tungsten wire with a fixed end dependence of frequency on temperature was $10^{-5}K^{-1}$. Experimental results and estimates of wire heating of existing scanners show, that the wire heats up to a few hundred grades, which is enough for measurements. Usage of wires of μm thickness diminishes the problem of the wire thermalisation speed during the scanning of the bunch.

## Introduction

Method of beam transverse profile measurement in accelerators by scanning wire is wide-spread in accelerator technology [1-7]. In such way measurement of transverse profiles of proton, antiproton, ion, electron, positron beams in wide range of energies is carried out without destruction of the beam [1, 3]. This method will also be used in new extrahigh energy accelerators (see, e.g. [7-9]).

The essence of the method is measurement of radiation and/or secondary particles arising at the scattering of the bean on scanning wire. Thus, the device consists of two essential units: a scanning mechanism with fixed on a fork wire and a system of radiation (particles) receiver, located along the beam propagation.

In this work a wire vibrating in its normal frequency is proposed to use as a wire scanner. Note, that pickups on the basis of vibrating wire are also wide-spread in technology and are used, in particular, in tensometry, gravimetry, measurements of magnetic field and magnetic properties of materials (see, e.g. [10-12]).

A beam of charged particles affects on vibrating wire frequency in several ways, namely: mechanical transverse influence caused by transferred pulse of the beam, influence of beam magnetic field, radiation affect of the beam on the wire material, heating of the wire. The last effect turned to be dominating, since for typical currents of accelerator beams the wire of μm diameter heats over 1000 K [3, 4].

In this work the strong dependence of normal frequency on its temperature and hence on the beam scattering is proposed to use for measurement of beam local intensity. Indubitable advantage of this method is the compactness of the whole system and the elimination of the unit of radiation receivers, which leads to the reduction of the price.

Authors of present work have accumulated a certain experience in development and producing of tension gauges on the basis of vibrating wire. In particular, good results on frequency long-time relative stability $\sim 10^{-5}$ at relative solution $\sim 10^{-6}$ are obtained. As result of frequency multiplication its measurement become much more accurate. We have used such a pickup for measurement of spatial distribution of magnetic filed [13-14].

Bellow we describe the main processes taking place at beam scattering on wire, results of wire heating modelling are presented, and specific problems, arising at development of vibrating wire scanner are discussed.

## Equilibrium temperature and wire thermalisation.

Consider a long thin round wire, strained long the $y$ axis in vacuum. Wire ends are kept at constant temperature $T_0$. "Thin" and "long" mean that the wire diameter $d$, its length $l$ and part of the wire irradiated by the beam $\sigma_y$ satisfy the condition $d \ll \sigma_y \ll l$. Beam propagates along the $z$ axis, the wire scans the beam in transverse direction along the $x$ axis.

Equilibrium temperature of an immobile wire in vacuum under beam is defined by the balance between the power $\Delta E/\Delta t$ brought to it by the beam and two factors of heat removal: thermal radiation through the cross section surface $W_1$ and thermoconductivity of the wire material through its ends $W_2$.

The power released on the wire can be estimated by formula:

$$\frac{\Delta E}{\Delta t} = k\left(\frac{dE}{dz}d\right)\left(p(x)\sigma_y d\right), \qquad (1)$$

where $dE/dz$ is the ionisation loss of beam particles in wire material, $p(x)$ is local flux density of beam particles, $k$ is the coefficient characterising the part of scattered beam energy working on wire heating.

The formula (1) has s simple meaning: the expression in first pair of brackets defines the losses of a particle when it passes through the wire material, the second member defines the number of scattered in a unit time particles. In works [3, 4] $k$ is set equal to 1/3.

The thermal radiation $W_1$ is determined by the formula:

$$W_1 = S_{side}\sigma(T^4 - T_0^4), \qquad (2)$$

where $S_{side}$ is the side surface of the wire ($S_{side} = \pi d\sigma_y$), $\sigma$ is the constant of Stefan-Boltzmann, $T$ is the absolute temperature of the wire.

Teat loss through the wire ends approximately is equal

$$W_2 = S\frac{2K}{l}(T - T_0), \qquad (3)$$

where $S = \pi\, d^2/4$ is the wire cross section, $K$ is the thermal conductivity of wire material, which depends on temperature.

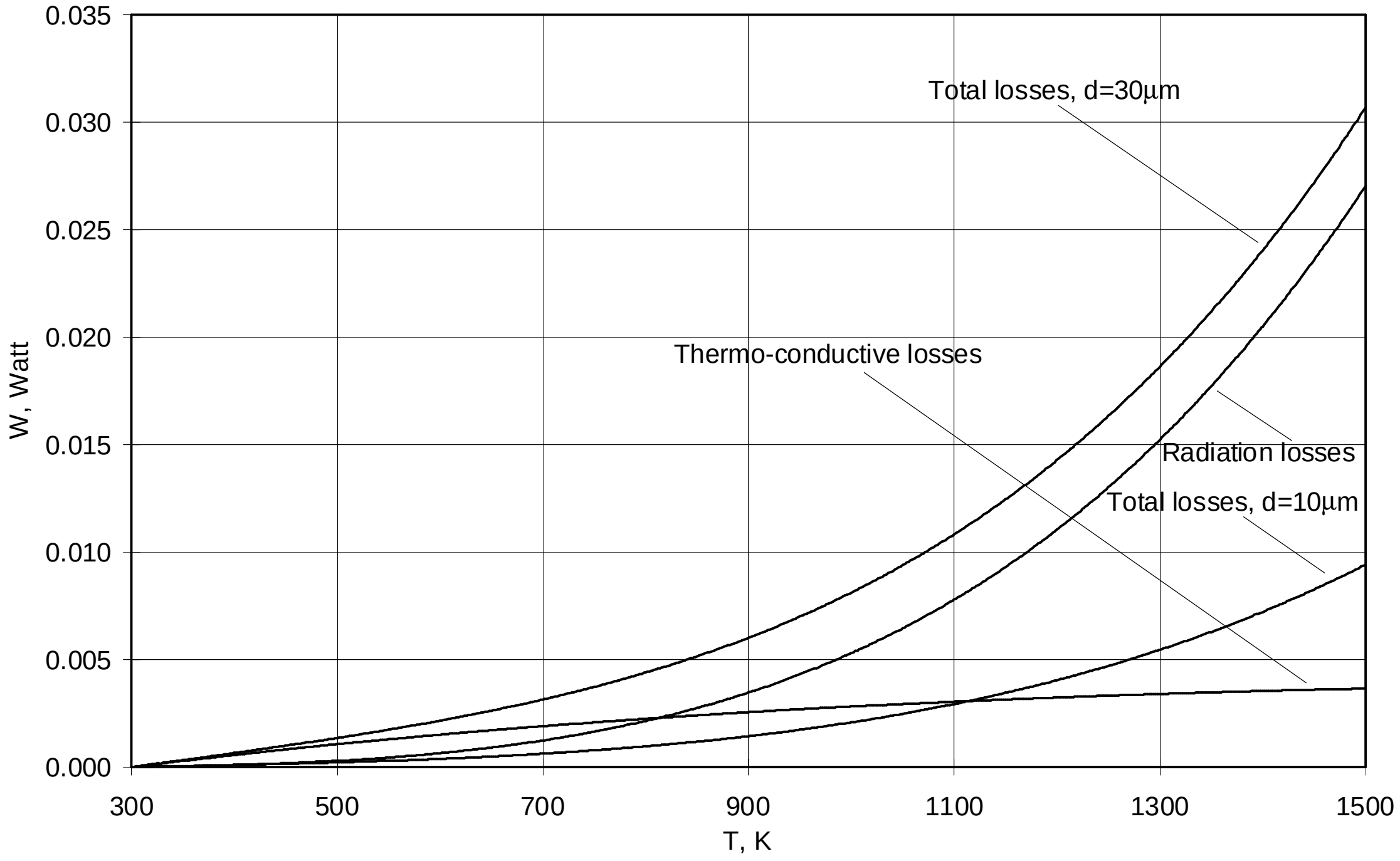

Fig. 1. Energy losses for tungsten wire, $d_1$=30 µm, $d_2$=10 µm, l=20 mm, $\sigma_x$=3 mm, $\sigma_z$=1 mm.

Fig. 1 shows calculated curves $W_1$ and $W_2$, also $W_1 + W_2$ for a tungsten wires of length $l$ = 40 mm and diameters $d_1$=10 µm and $d_2$=30 µm ($\sigma_y$ =1 mm). One can see that at the same absorption power the thin wire is heated more than the thick one, and this difference rises with increment of power. Presented curves show that to heat the wire of diameter 30 µm up to 1000K power ≈10 mW is needed. Taking into account that the wire resistance ≤ 1 Ω one can conclude that current of the order of few hundred mA is needed.

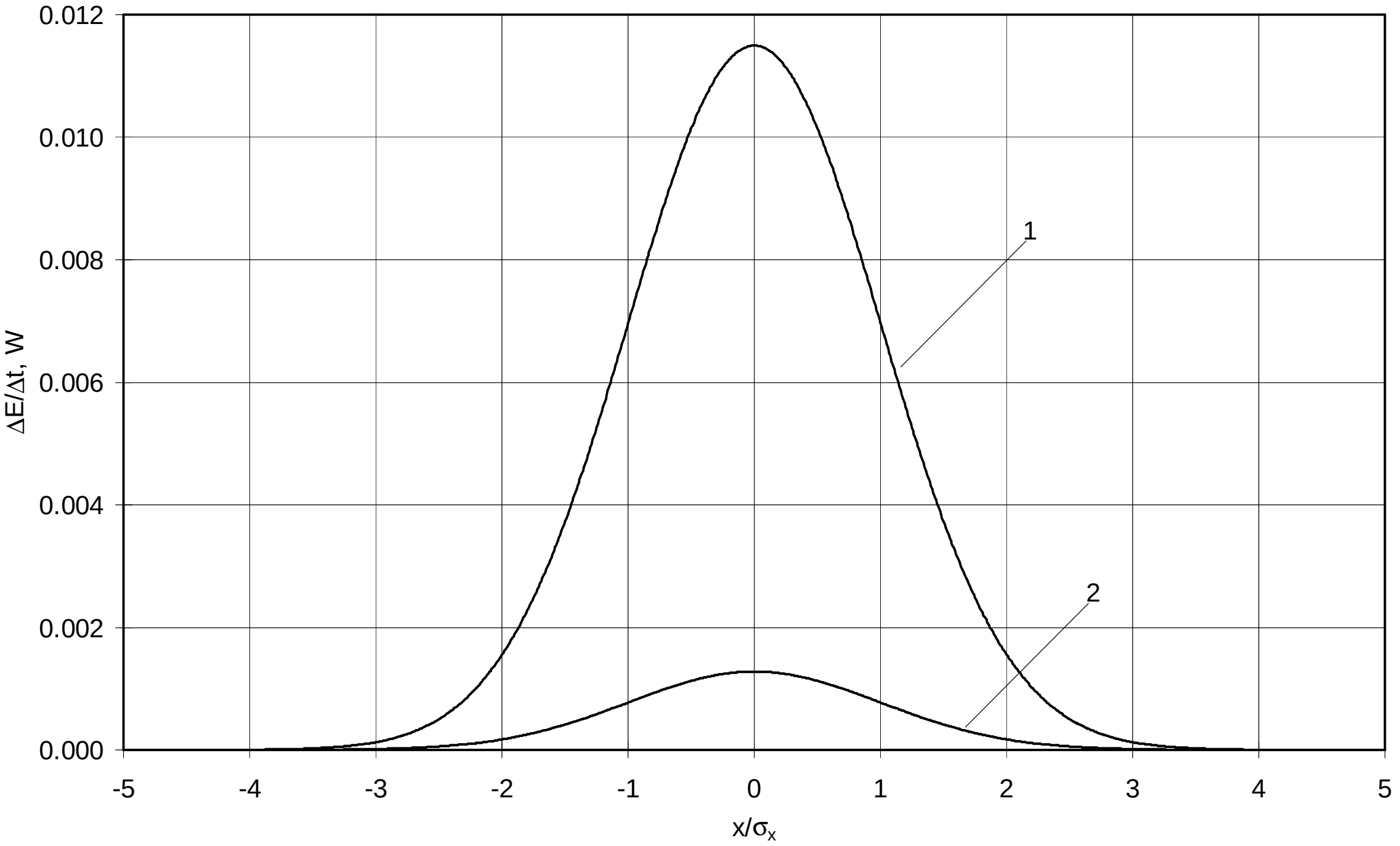

Fig. 2. Heat energy from bunch scattering, I=0.9 mA, $d_1$=30 μm, $d_2$=10 μm.

Fig. 2 shows the values of energy released on wires of diameters 10 μm and 30 μm at scattering on it of a beam with average current 0.9 mA depending on relative distance of the wire from centre of the beam with horizontal transverse size $\sigma_x$ = 3 mm, calculated by the formula (1). When calculating the value of ionisation losses $\Delta E/\Delta t$ is set equal to 5.76 J/m [5]. The current distribution was supposed Gaussian in beam cross section. One can see from presented curves, that at the distances more than $3\sigma_x$ one can neglect the wire heating, and rather abrupt dependence of separated energy is at distances up to $3\sigma_x$ from beam centre. Characteristic values of beam energy released on the wire are few mWs.

The equilibrium temperature of the wire is determined from the equation of balance:

$$W_1 + W_2 = \Delta E/\Delta t\,. \tag{4}$$

Solution of this equation with respect to $T$ at different distances of the wire from the beam centre are presented in Fig. 3. One can see that for wires of diameters 10 μm è 30 μm the maximum temperatures are reached when the wire is close to the beam centre and are equal to 1100 K and 850 K respectively.

Time of thermalisation $\tau$ also is an important parameter for the scanning. To estimate this time one can use the formula

$$\tau \cong E_{store}/W\,, \tag{5}$$

where $E_{store}$ is the thermal energy stored in the wire, ($E_{store} = (\pi d^2 / 4)\sigma_y c_V (T - T_0)$, where $c_V$ - is the heat capacity per unit volume of the wire material, for tungsten $c_V = 2.86 \times 10^6 \mathrm{Jm^{-3}K^{-1}}$). Fig. 4 represents calculated values of thermalisation times for tungsten wires of diameters 10 μm and 30 μm. As it is seen from graphics, the thermalisation time is minimal in the beam centre and is 2 and 3 seconds respectively. In the beam periphery (at the distances $2\sigma_x$) this time is approximately doubled.

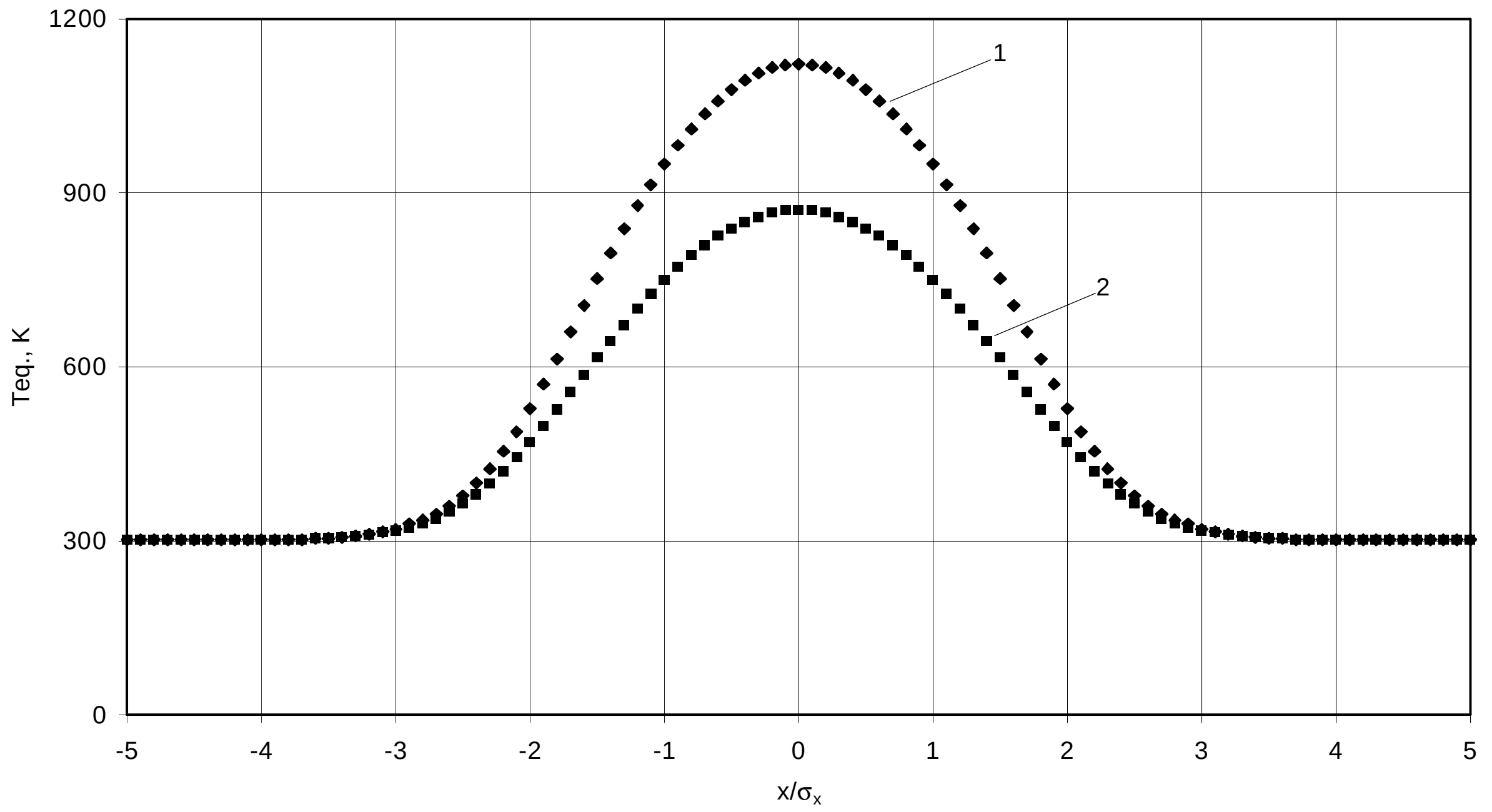

Fig. 3. Wire equilibrium temperature, $d_1$=30 μm, $d_2$=10 μm.

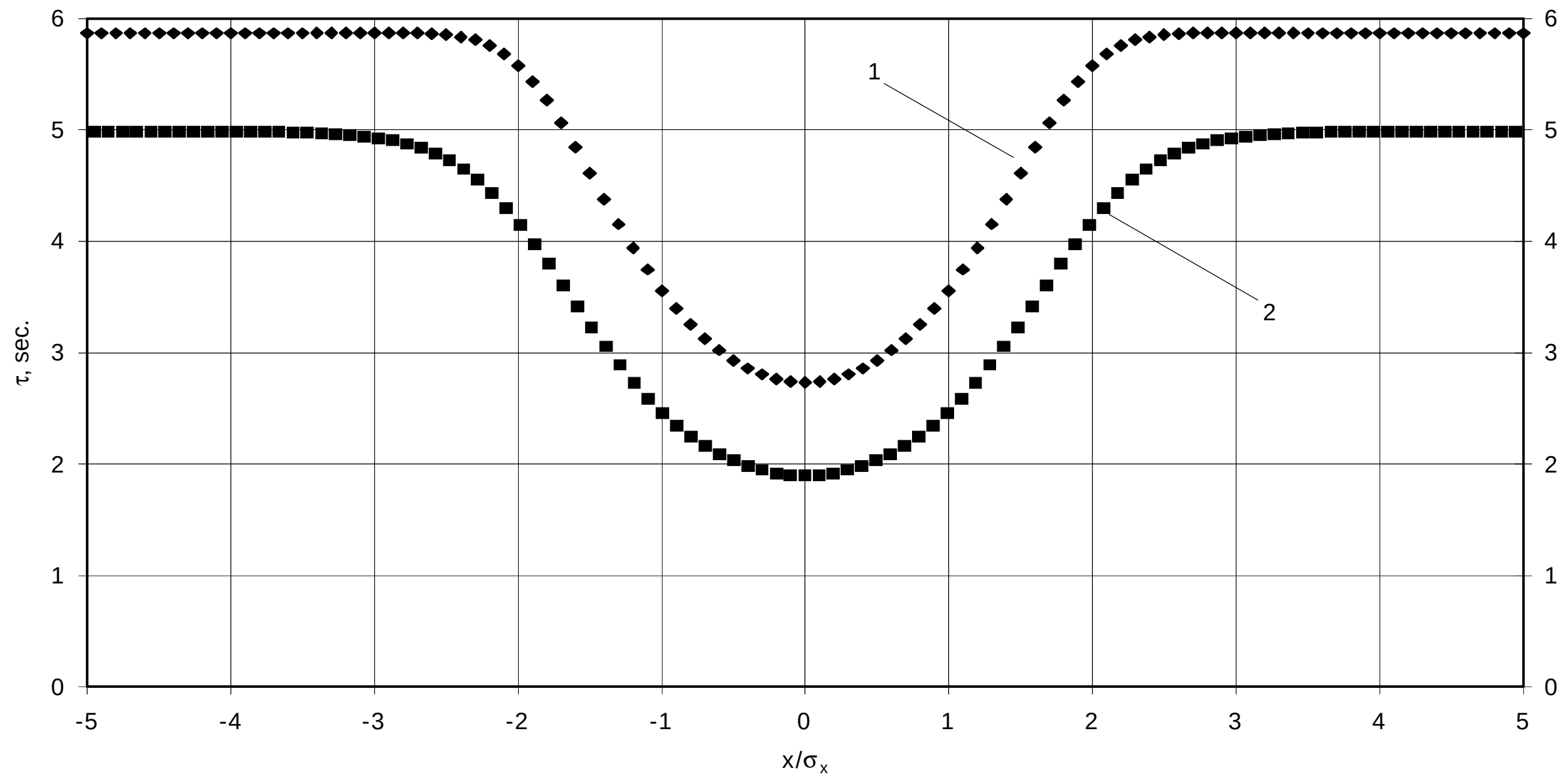

Fig. 4. Wire thermalisation time, $d_1$=30 μm, $d_2$=10 μm.

Note when scanning consecutively the total time of scanning is much more less than the sum of thermalisation times of individual points, because the wire temperature changes insignificantly at passing from one point to another.

Estimate of transverse force for relativistic beams acting on wire gives values of the order of $W/c$ ~$10^{-11}$ N. Thus, one can neglect the transverse force affect frequency.

Model Experiment

As it was shown above the main influence of the beam on wire is its heating. Influence of the wire temperature on its normal oscillations frequency is determined by its elongation:

$$\Delta f/f = -(1/2)\,\Delta l/l = \alpha_b \Delta T \,, \tag{6}$$

Where $\Delta T$ is the temperature variation. Coefficient $\alpha_b$ for metals varies from few units of $10^{-6}$ $K^{-1}$ (tungsten, molybdenum) up to $10^{-5}$ $K^{-1}$ (steels, bronzes, brasses, aluminium alloys). Some special alloys have extralow temperature coefficient of thermal expansion, e.g. for invar $\alpha$ =(0.01-2.0)$\times 10^{-6}$ $K^{-1}$ [15].

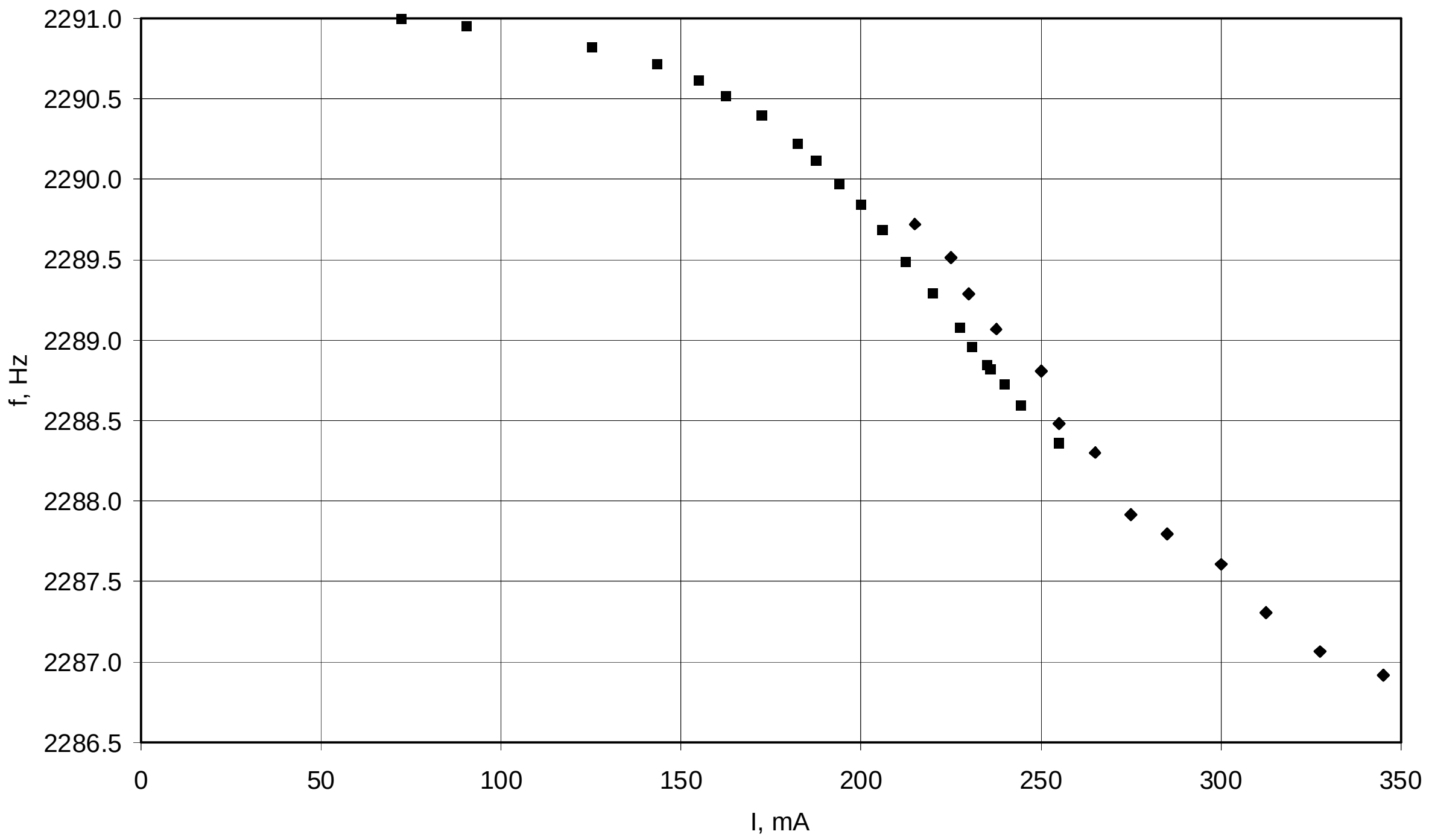

Fig. 5. Frequency dependence on current.

However, for the rigidly fixed wire more essential is the influence of temperature changes on wire tension arising from inequality of coefficients of thermal expansion of wire $\alpha_s$ and the base $\alpha_b$. Thus [16]:

$$\frac{\Delta f}{f} = \frac{1}{2}\frac{\Delta F}{F} \approx \frac{ES}{F}(\alpha_b - \alpha_s)\Delta T, \tag{7}$$

where $E$ - is the Young's modulus, $S$ is the wire cross section. At deducing (7) it is taken into account that $\Delta F/F = E\,\Delta l/l$. In formula (7) a large dimensionless factor $ES/F$ is separated out. For the tungsten wires of diameter ~100 μm the upper limit of loads for stable frequency generation is less than 10 N. In this case $ES/F > 400$. For the beryl bronze wire of diameter 90μm and $F$ ~3 N near the upper limit of operating range of loads $ES/F$~300. Thus, for the rigidly fixed wire the dominating factor is the influence of the temperature on wire tension.

As it was shown above the main result of the irradiation of the wire by the beam is its heating. Heating of the scanning wire was modelled by alternating and direct currents through the wire. Wire normal oscillations were generated in two modes: with one free end and with fixed ends of the wire. The experiments were done in forevacuum evacuation.

In experiments with alternating current forced oscillations of current frequency arises (at the current frequency 50 Hz normal oscillations broke at the current $I$ > 50 mA).

When heating by direct current the same tungsten wires of diameter 70 μm were used.

Data obtained at constant wire tension are presented in Fig. 5. Results of two series of measurements for are presented. The first series (squares) was carried out for currents from 75mA up to 250 mA; glow appeared at 235 mA. Later on an experiment with currents from 215mA up to 345 mA was done (diamonds). In first series the current was changed stepwise, increasing consequently from the lower values to the upper ones. In second case between two consequent values the current was nullified, each current value being held during 2 minutes. In second series the frequency had restored well at each switching on for currents up to 345 mA. At upper values of the current after each heating cycle the frequency systematically went up. Yellow glow ($T \approx 1000$ Ê) appears currents more than 300 mA.

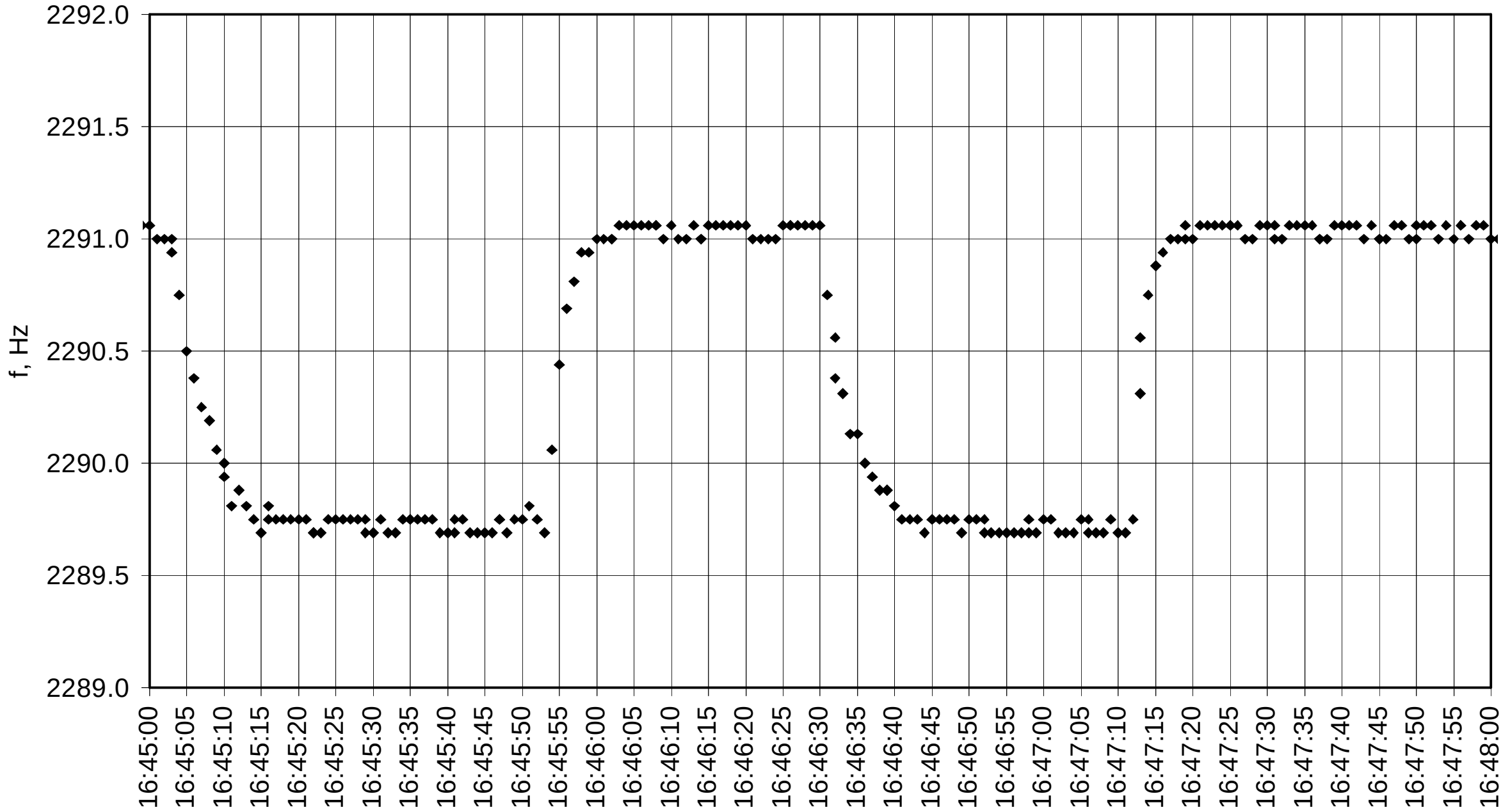

Fig. 6 Transient characteristics of wire with a free end, I=215mA.

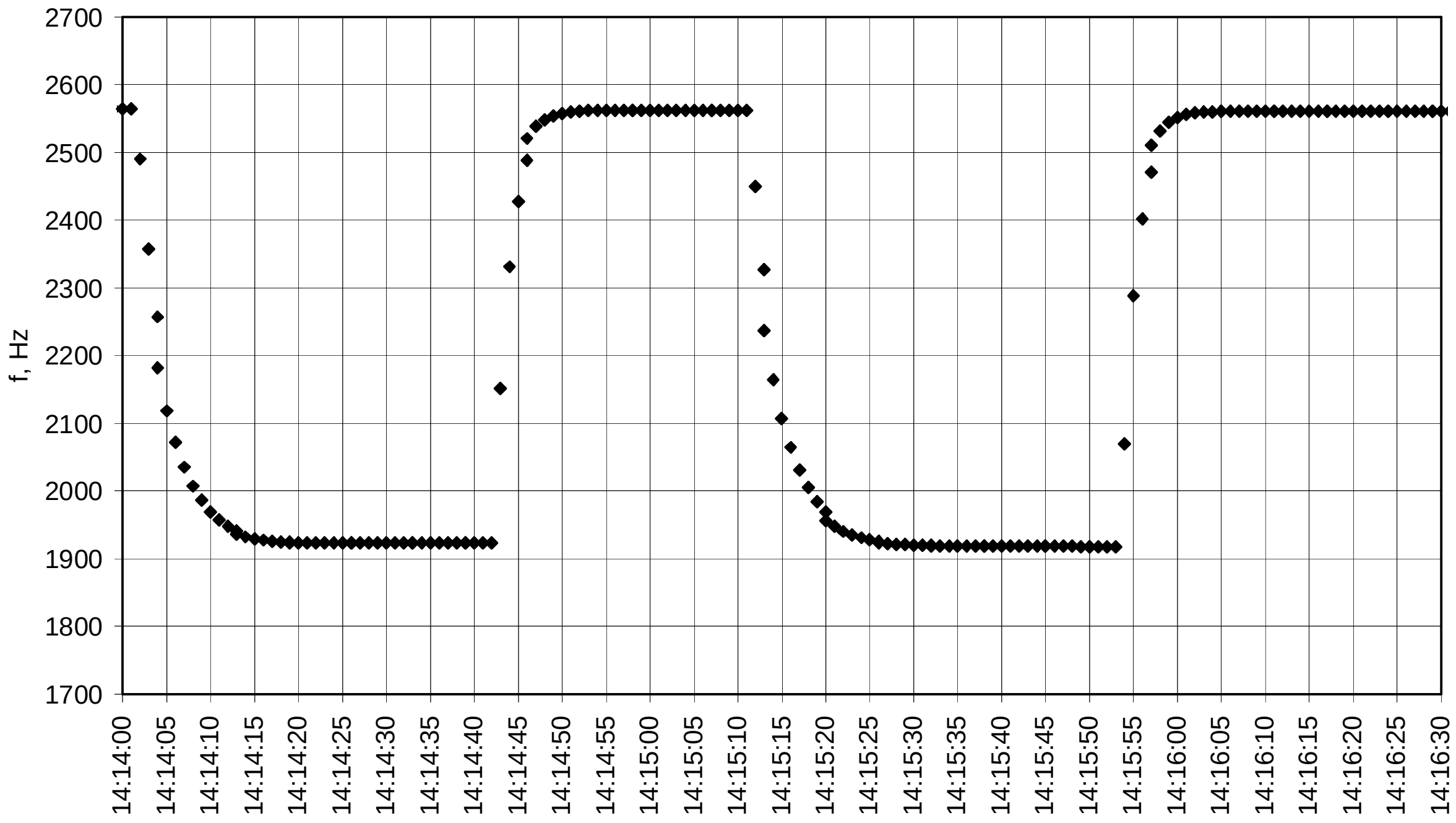

Fig. 7. Transient characteristics of wire with fixed ends, I=215mA.

Fig. 6 shows the transient characteristics at switching on/off for current 215 mA. One can see that the frequency exceeds its steady-state value during a few seconds. One can take this value for estimate of the thermalisation time $\tau$ . Cooling down processes are depicted without destruction, because after switching on it takes some time for current to reach its steady-state.

Similar experiments were done for wires with fixed ends. As it was mentioned above, the frequency changes interval was much more wider. Fig. 7 shows the transient characteristics at switching on/off of the current 215 mA.

Restoration of the frequency in both cases (Figs. 6 and 7) were good enough, i.e. heating of the wire up to 1000 K does not lead to irreversible changes of wire parameters.

Thus, the model experiments show, that reproducible generation of wire normal modes is possible at its considerable heating. And theoretical estimates well accord with obtained experimental data.

<u>Wire under Bunch - Proposal for Experiment</u>

Let's discuss some specific problems, which can arise at scanning of the accelerator bunch by a vibrating wire.

*Magnetic field*

There are two possibilities in scheme of wire scanner: to use a scheme with own magnet or use the accelerator magnetic fields (dipole magnets, quadrupole lenses).

In developed by the authors pickups normal oscillations of the wire were excited in autogeneration scheme. Exciting action on the wire arises as a result of interaction between the current through the wire and magnetic field of the samarium cobalt magnet, the field being localised in an magnet operating gap of the length 10 mm. The characteristic field inside the gap was about 8 kGs.

To keep the transverse effective square of the wire with respect to on-going bunch, separation of oscillations along the bunch axis is desirable. Accelerator magnetic fields meet this requirement. It can also be satisfied at own magnet using.

Fields of dipole magnets are strong enough, however, they fix the horizontal movement of the wire.

It is also possible to use the fields of quadrupole lenses. However, in such fields even modes of the wire normal oscillations are possible. Scanning can be done in vertical, horizontal or under the angle $45^0$ directions.

Usage of vibrating wire scanners with own magnets located in accelerators free spaces is also possible. In so doing it is preferable to intersect the bunch at 1/4 of wire length, the magnetic circuit being located symmetrically with respect to wire centre, in such way providing generation of the second harmonic. In this case the magnetic field outside the gap is to be carefully screened.

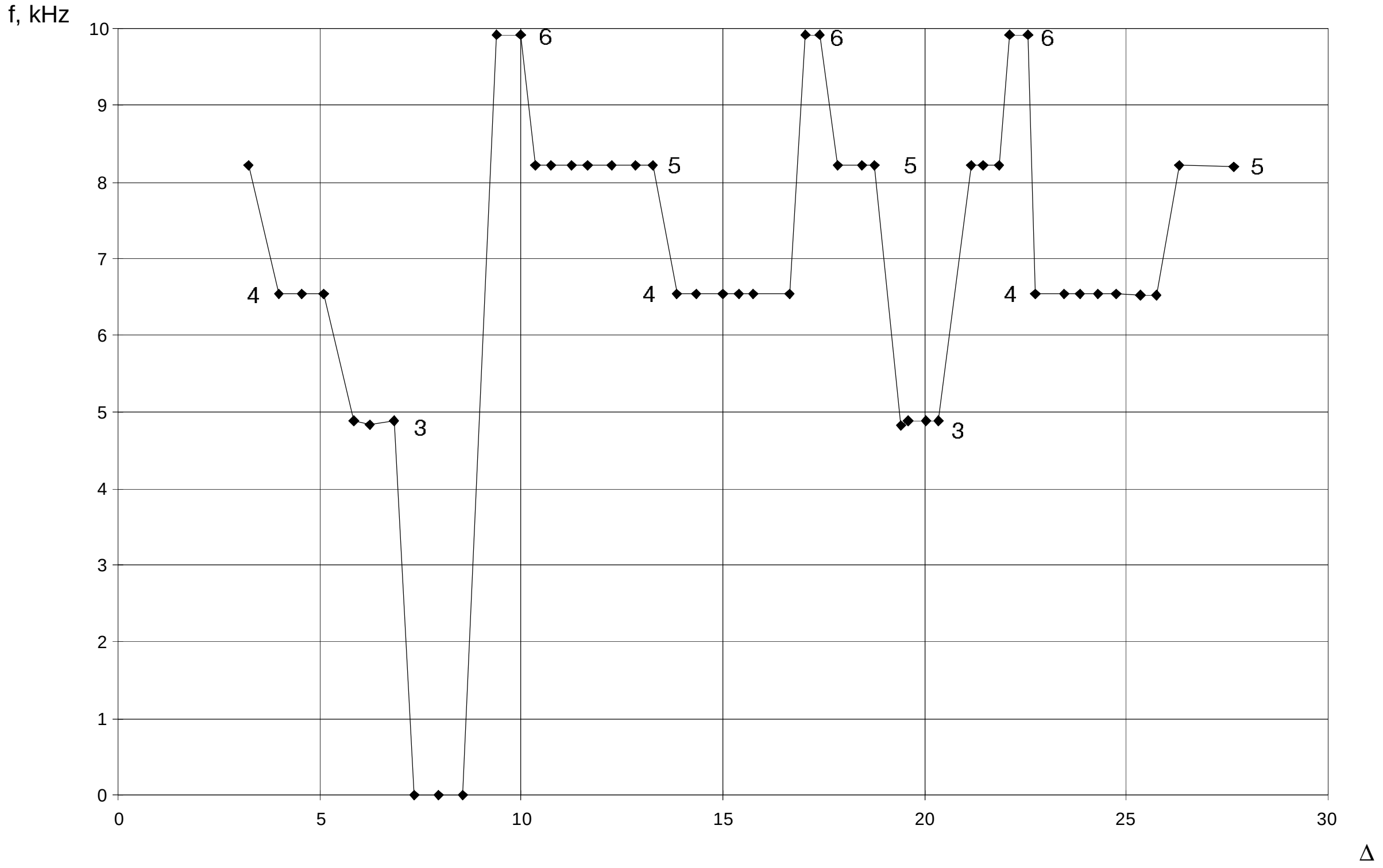

Fig. 8. Higher harmonics generation depending on magnet shift from upper end of wire.

To demonstrate the possibility of generation of different harmonics we have carried out an experiment with shortened magnetic circuit, shifted along the wire. Results of experiment are presented in Fig. 8. The horizontal axis is the shift of the centre of magnetic circuit with respect two fixed end of the wire. Position 20 corresponds to the wire middle. Shift of the magnetic circuit downwards (in Fig. 8 to right) was limited by the construction, however, one can see that excitation of normal harmonics of the wire is symmetric with respect to its middle. Usage of short magnetic circuit results in generation of third harmonic in central position. By shifting of the magnetic circuit one can obtain generation of third, fourth, fifth or sixth harmonics (in figure are indicated by numbers). There was a silence zone, too. Of course, experiments results depend on geometric characteristics, however, they show the possibility of stable generation of needed harmonic.

Note one can neglect the self magnetic fields of the bunch. These fields are much more less than the external accelerator fields, and, secondly their affection results in a small influence of magnetic field amplitude (wire oscillations amplitude, indeed) on frequency.

*Radiation Resistance*

Tungsten and bronze (beryl bronze) are used both in scanning wire and vibrating wire technologies. In both cases they show high radiation resistance. The electronic circuit exciting the wire scanner oscillations is located far from the wire and can be sufficiently screened from radiation. Other parts of vibrating wire scanner are passive elements and are also radiation resistant.

*Thermal Stability, Thermalisation Times*

Stability with respect to wire heating is one of the main problems in developed wire scanner technologies. To solve this problem, the scanning was done at a high speed, up to ≈10m/sec. [3, 4], i.e. the wire passes across the mm size bunch during $10^{-4}$ seconds. Such times are too small for thermalisation of the wire. Note, the estimate (5) of $\tau$ can be lowered, because in real movement from one point to another will result in less temperature jumps. However, one cannot expect thermalisation times less than $10^{-2}$ at temperatures, where long time exploitation of wire is possible.

Usage of thin and short wires will give some advantage.

It is possible that limit on thermalisation time can be solved by improvement of wire fixing system, which now is essentially simplified in comparison with similar wire scanning systems.

Mainly peripheral monitoring of the bunch and its accompanying radiation also can be interesting.

*Vacuum System*

Since the vibrating scanner system consist of one unit, its installation into accelerator vacuum system will accompany less problems than in the case of two-unit wire scanner system. It is also desired to have a system of wire replacing without loss of vacuum in acceleration chamber.

*Frequency measurement*

The problem frequency quick measurement is solved by its preliminary multiplication by a large factor. Thus, in experimentally realised pickups multiplication by 32 allows to measure few kHz frequencies with accuracy 0.03 Hz during 1 sec.

## Conclusion

Indubitable advantage of proposed method is the compactness of the whole system and the elimination of the unit of radiation receivers. The peripheral monitoring of the bunch by proposed method also can be interesting.

Authors would like to thank R.Reetz and A.Ts.Amatuni for helping support.

## References

1. Suwada T. et al. First beam test result of a prototype wire scanner for the KEKB injector linac and BT lines. - KEK Preprint 97-184, 1997.
2. Martini M. et al. Emittance measurements of the CERN PS complex.- CERN-PS-97-018-CA.
3. Steinbach Ch. Emittance measurements with the CERN PS wire scanner. - CERN/PS 95-04(OP).
4. Agoritsas V. et al. The first wire scanner of the CERN PS, - CERN/PS 95-06 (BD/OP).
5. Bosser J. et al. NIM, A234, 475 (1985).
6. Tatchan R. Analysis a novel diffractive scanning-wire beam position monitor (BPM) for discriminative profiling of electon vs. X-ray beams.- PAC'98, p.2192.
7. Fulton R. et al. A high resolution wire scanner for micron-size profile measurements of the SLC. - SLAC-PUB-4605, 1988.
8. Field C. The wire scanner system of the Final Focus Test Beam. - SLAC-PUB-6717, 1994.
9. McCormick D. et al. Measuring micron size beams in the SLC Final Focus, - SLAC-PUB-6615, 1994.
10. Methods of Experimental Physics, Vol.1, Classical Methods, Academic Press, New York and London, 1959.
11. El'tsev Ju. F., Zakosarenko V.M., Tsebro V.I., String Magnetometer, Trudy FIAN, v. 150, M, 1984.
12. Billan J., Materials, in CERN Accelerator School, Magnet measurement and alignment, (Switzerland, 16-20 March 1992), Proceedings, Ed. S. Turner CERN 92-05, p. 17.
13. Arutunian S.G., et al, Gravimetric measurement of magnetic field gradient spatial distribution.- Pribory I technika experimenta (in press).
14. Arutunian S.G. et al, Magnetic field distribution measurement by vibrating wire strain gauge.- PAC'99.
15. I.S. Grigorjev, E.Z. Mejlikhov, Physical Data, Handbook.- M., Energoatomizdat, 1991.
16. Agejkin D.I. et al. Control and Regulation Pickups.- M., Mashinostroenie, 1965.